\documentstyle[twoside,psfig]{article}

\catcode`\@=11
\long\def\@makefntext#1{
\protect\noindent \hbox to 3.2pt {\hskip-.9pt  
$^{{\eightrm\@thefnmark}}$\hfil}#1\hfill}		

\def\thefootnote{\fnsymbol{footnote}}
\def\@makefnmark{\hbox to 0pt{$^{\@thefnmark}$\hss}}	
	
\def\ps@myheadings{\let\@mkboth\@gobbletwo
\def\@oddhead{\hbox{}
\rightmark\hfil\eightrm\thepage}   
\def\@oddfoot{}\def\@evenhead{\eightrm\thepage\hfil
\leftmark\hbox{}}\def\@evenfoot{}
\def\sectionmark##1{}\def\subsectionmark##1{}}

\oddsidemargin=\evensidemargin
\renewcommand{\thefootnote}{\fnsymbol{footnote}}

\newcounter{sectionc}\newcounter{subsectionc}\newcounter{subsubsectionc}
\renewcommand{\section}[1] {\vspace{12pt}\addtocounter{sectionc}{1} 
\setcounter{subsectionc}{0}\setcounter{subsubsectionc}{0}\noindent 
	{\tenbf\thesectionc. #1}\par\vspace{5pt}}
\renewcommand{\subsection}[1] {\vspace{12pt}\addtocounter{subsectionc}{1} 
	\setcounter{subsubsectionc}{0}\noindent 
	{\bf\thesectionc.\thesubsectionc. {\kern1pt \bfit #1}}\par\vspace{5pt}}
\renewcommand{\subsubsection}[1] {\vspace{12pt}\addtocounter{subsubsectionc}{1}
	\noindent{\tenrm\thesectionc.\thesubsectionc.\thesubsubsectionc.
	{\kern1pt \tenit #1}}\par\vspace{5pt}}
\newcommand{\nonumsection}[1] {\vspace{12pt}\noindent{\tenbf #1}
	\par\vspace{5pt}}

\newcounter{appendixc}
\newcounter{subappendixc}[appendixc]
\newcounter{subsubappendixc}[subappendixc]
\renewcommand{\thesubappendixc}{\Alph{appendixc}.\arabic{subappendixc}}
\renewcommand{\thesubsubappendixc}
	{\Alph{appendixc}.\arabic{subappendixc}.\arabic{subsubappendixc}}

\renewcommand{\appendix}[1] {\vspace{12pt}
        \refstepcounter{appendixc}
        \setcounter{figure}{0}
        \setcounter{table}{0}
        \setcounter{lemma}{0}
        \setcounter{theorem}{0}
        \setcounter{corollary}{0}
        \setcounter{definition}{0}
        \setcounter{equation}{0}
        \renewcommand{\thefigure}{\Alph{appendixc}.\arabic{figure}}
        \renewcommand{\thetable}{\Alph{appendixc}.\arabic{table}}
        \renewcommand{\theappendixc}{\Alph{appendixc}}
        \renewcommand{\thelemma}{\Alph{appendixc}.\arabic{lemma}}
        \renewcommand{\thetheorem}{\Alph{appendixc}.\arabic{theorem}}
        \renewcommand{\thedefinition}{\Alph{appendixc}.\arabic{definition}}
        \renewcommand{\thecorollary}{\Alph{appendixc}.\arabic{corollary}}
        \renewcommand{\theequation}{\Alph{appendixc}.\arabic{equation}}
        \noindent{\tenbf Appendix \theappendixc #1}\par\vspace{5pt}}
\newcommand{\subappendix}[1] {\vspace{12pt}
        \refstepcounter{subappendixc}
        \noindent{\bf Appendix \thesubappendixc. {\kern1pt \bfit #1}}
	\par\vspace{5pt}}
\newcommand{\subsubappendix}[1] {\vspace{12pt}
        \refstepcounter{subsubappendixc}
        \noindent{\rm Appendix \thesubsubappendixc. {\kern1pt \tenit #1}}
	\par\vspace{5pt}}

\newcommand{\textlineskip}{\baselineskip=13pt}
\newcommand{\smalllineskip}{\baselineskip=10pt}

\def\eightcirc{
\begin{picture}(0,0)
\put(4.4,1.8){\circle{6.5}}
\end{picture}}
\def\eightcopyright{\eightcirc\kern2.7pt\hbox{\eightrm c}} 

\newcommand{\copyrightheading}[1]
	{\vspace*{-2.5cm}\smalllineskip{\flushleft
	{\footnotesize Modern Physics Letters A, #1}\\
	{\footnotesize $\eightcopyright$\, World Scientific Publishing
	 Company}\\
	 }}

\newcommand{\publisher}[2]{{\begin{center}\footnotesize\smalllineskip 
	Received #1\\
	Revised #2
	\end{center}
	}}

\def\abstracts#1#2#3{{
	\centering{\begin{minipage}{4.5in}\footnotesize\baselineskip=10pt
	\parindent=0pt #1\par 
	\parindent=15pt #2\par
	\parindent=15pt #3
	\end{minipage}}\par}} 

\newcommand{\bibit}{\nineit}
\newcommand{\bibbf}{\ninebf}
\renewenvironment{thebibliography}[1]
	{\frenchspacing
	 \ninerm\baselineskip=11pt
	 \begin{list}{\arabic{enumi}.}
        {\usecounter{enumi}\setlength{\parsep}{0pt}     
	 \setlength{\leftmargin 12.7pt}{\rightmargin 0pt} 
         \setlength{\itemsep}{0pt} \settowidth
	{\labelwidth}{#1.}\sloppy}}{\end{list}}

\newcounter{itemlistc}
\newcounter{romanlistc}
\newcounter{alphlistc}
\newcounter{arabiclistc}

\newcommand{\fcaption}[1]{
        \refstepcounter{figure}
        \setbox\@tempboxa = \hbox{\footnotesize Fig.~\thefigure. #1}
        \ifdim \wd\@tempboxa > 5in
           {\begin{center}
        \parbox{5in}{\footnotesize\smalllineskip Fig.~\thefigure. #1}
            \end{center}}
        \else
             {\begin{center}
             {\footnotesize Fig.~\thefigure. #1}
              \end{center}}
        \fi}

\newcommand{\tcaption}[1]{
        \refstepcounter{table}
        \setbox\@tempboxa = \hbox{\footnotesize Table~\thetable. #1}
        \ifdim \wd\@tempboxa > 5in
           {\begin{center}
        \parbox{5in}{\footnotesize\smalllineskip Table~\thetable. #1}
            \end{center}}
        \else
             {\begin{center}
             {\footnotesize Table~\thetable. #1}
              \end{center}}
        \fi}

\def\@citex[#1]#2{\if@filesw\immediate\write\@auxout
	{\string\citation{#2}}\fi
\def\@citea{}\@cite{\@for\@citeb:=#2\do
	{\@citea\def\@citea{,}\@ifundefined
	{b@\@citeb}{{\bf ?}\@warning
	{Citation `\@citeb' on page \thepage \space undefined}}
	{\csname b@\@citeb\endcsname}}}{#1}}

\newif\if@cghi
\def\cite{\@cghitrue\@ifnextchar [{\@tempswatrue
	\@citex}{\@tempswafalse\@citex[]}}
\def\citelow{\@cghifalse\@ifnextchar [{\@tempswatrue
	\@citex}{\@tempswafalse\@citex[]}}
\def\@cite#1#2{{$\null^{#1}$\if@tempswa\typeout
	{IJCGA warning: optional citation argument 
	ignored: `#2'} \fi}}

\def\pmb#1{\setbox0=\hbox{#1}
	\kern-.025em\copy0\kern-\wd0
	\kern.05em\copy0\kern-\wd0
	\kern-.025em\raise.0433em\box0}

\def\fnt#1#2{\footnotetext{\kern-.3em
	{$^{\mbox{\scriptsize #1}}$}{#2}}}

\def\fpage#1{\begingroup
\voffset=.3in
\thispagestyle{empty}\begin{table}[b]\centerline{\footnotesize #1}
	\end{table}\endgroup}

\def\runninghead#1#2{\pagestyle{myheadings}
\markboth{{\protect\footnotesize\it{\quad #1}}\hfill}
{\hfill{\protect\footnotesize\it{#2\quad}}}}
\font\tenrm=cmr10
\font\tenit=cmti10 
\font\tenbf=cmbx10
\font\bfit=cmbxti10 at 10pt
\font\ninerm=cmr9
\font\nineit=cmti9
\font\ninebf=cmbx9
\font\eightrm=cmr8

\def\qed{\hbox{${\vcenter{\vbox{			
   \hrule height 0.4pt\hbox{\vrule width 0.4pt height 6pt
   \kern5pt\vrule width 0.4pt}\hrule height 0.4pt}}}$}}

\renewcommand{\thefootnote}{\fnsymbol{footnote}}	
\def\newline{\hfill\break}

\def\mod{\;{\rm mod}\;}
\def\tr{\;{\rm tr}\;}

\def\ignore#1{\ }

\newcommand{\grad}{\nabla}
\newcommand{\curl}{\nabla \times}
\newcommand{\pdr}{\partial}
\newcommand{\eps}{\epsilon}
\newcommand{\half}{{1 \over{2}}}

\newcommand{\beq}{\begin{equation}}
\newcommand{\eeq}{\end{equation}}
\newcommand{\bea}{\begin{eqnarray}}
\newcommand{\eea}{\end{eqnarray}}
\def\m#1{$#1$}
\def\mbf#1{{\mbox{\boldmath $#1$}}}
\begin{document}
\setlength{\textheight}{7.7truein}  

\runninghead{ Two Dimensional Turbulence }
{Using Random Matrix Theory}
\normalsize\textlineskip
\thispagestyle{empty}
\setcounter{page}{1}

\copyrightheading{}			

\vspace*{0.88truein}

\fpage{1}
\centerline{\bf A Model of Two Dimensional Turbulence Using }
\baselineskip=13pt
\centerline{\bf Random Matrix Theory}
\vspace*{0.37truein}

\centerline{\footnotesize Savitri V. Iyer}
\baselineskip=12pt
\centerline{\footnotesize\it State University of New York at Geneseo}
\baselineskip=10pt
\centerline{\footnotesize\it Geneseo, NY 14454, USA}
\vspace*{0.225truein}

\centerline{\footnotesize S.G. Rajeev}
\baselineskip=12pt
\centerline{\footnotesize\it University of Rochester}
\baselineskip=10pt
\centerline{\footnotesize\it Rochester, NY 14627, USA}
\vspace*{10pt}

\publisher{(received date)}{(revised date)}

\vspace*{0.21truein}
\abstracts{We derive a formula for the entropy of two dimensional
incompressible inviscid flow, by determining the volume of the space
of vorticity distributions with fixed values for the moments \m{Q_k=\int
\omega(x)^k d^2x}. This space is approximated by a sequence of 
 spaces of finite volume, by using a regularization of  the
 system that is geometrically natural and connected with the theory of random
 matrices. In taking the limit we get a simple
 formula for the entropy of a vortex field. We predict vorticity
 distributions of maximum entropy with given mean vorticity and enstrophy;
 also we predict the cylindrically symmetric vortex field with maximum
 entropy. This could be an approximate description of a hurricane.}{}{}



\vspace*{1pt}\textlineskip	
\section{Introduction}	
\vspace*{-0.5pt}
\noindent

Two dimensional inviscid incompressible flow is completely determined by
solutions to the Euler equations in the absence of random external forces.
Detailed numerical  as well as analogue simulations\cite{dezhe} show complex phenomena
such as recombination of vortices. Still, the situation is much simpler than in
three dimensions. It appears that energy flows from small scales to large scales
so that over  long time intervals the flow becomes smoother. This is the
opposite of the behavior observed in three dimensional hydrodynamics. Thus two
dimensional flow is not turbulent in the same sense as three dimensional flow.
Nevertheless, there is sufficient complexity remaining even in two dimensions
that a statistical approach is worthwile.

The study of the statistical mechanics of
inviscid flow dates back to Onsager\cite{onsager} with some recent revivals\cite{sommeria,jmiller}. The
crucial input into any statistical theory is the formula for entropy.
Previous studies have used various postulates for the entropy of a 
vortex field  \m{\omega(x)} such as 
\beq
	\int \omega(x)\log|\omega(x)|d^2x.
\eeq
We will determine the formula for entropy predicted by its microscopic
(Boltzmann) definition as the logarithm of the volume of the phase space
with a fixed value of the conserved quantities (more precisely, central
observables---see below). The conserved quantities of
the Euler equation are the moments \m{Q_k=\int \omega(x)^k d^2x}. The volume
of the phase space is difficult to determine directly because it is
infinite dimensional. 

We will determine a formula for entropy by `regularizing' the system; i.e., 
approximating
it with a finite dimensional system. The volume of the phase space is then
finite---indeed the answer is known in the literature on Random Matrix theory\cite{mehta}. The entropy of the original system can then be determined by taking
the limit as the dimension goes to infinity. We establish this way that the
formula for entropy is 
\beq
\chi={1\over A^2}\int \log|\omega(x)-\omega(y)|d^2x d^2y
\eeq 
where \m{A} is the area of the region within which the flow is contained.
This is quite different from the  postulates used in earlier analyses\cite{onsager,sommeria,jmiller}. We then use this formula to predict the maximum entropy
configuration with given mean vorticity and enstrophy\footnote{{\em Enstrophy}
is the second moment of vorticity: \m{Q_2=\int \omega^2(x)d^2x}.}
---it is the Wigner
semi-circular distribution. We  predict the vorticity
distribution of an axi-symmetric vortex of maximum entropy.

Polyakov\cite{polyakov}  has presented another theory of two dimensional turbulence, based on
the group of conformal transformations. Our considerations are in many ways
orthogonal, being based on the group of area preserving transformations rather
than  conformal transformations (which preserve instead the angles). 

In Section 2 we present an overview of two dimensional hydrodynamics.  The
Euler equations are formulated as a hamiltonian dynamical system, in
analogy to the rigid body equations also due to Euler.

In Section 3 we discuss the regularization procedure.  We use ideas that
have their origin in quantum field theory, but have already been used in
the hydrodynamics literature\cite{discrete}.

In Section 4 we reformulate the regularized system in terms of hermitean matrices 
which allows us to use the ideas from random matrix theory to derive a formula for entropy in Section 5. In Section 6 we derive the distribution function for vorticity that maximizes entropy for given mean vorticity and enstrophy. And finally we predict the axi-symetric vortex field of maximum entropy in Section 7.

\setcounter{footnote}{0}
\renewcommand{\thefootnote}{\alph{footnote}}

\section{Two-Dimensional Hydrodynamics as a Hamiltonian System}
\noindent

Incompressible  hydrodynamics is  described by  the Navier-Stokes equation:
\beq 
\quad {D{\bf u}\over Dt}=-\nabla p+R^{-1} \Delta {\bf u}+{\bf f} 
\eeq
along with the constraint $\nabla \cdot {\bf u}=0$.  Here, ${\bf u},p,R,{\bf
f}$ are  the velocity field, the pressure, the Reynolds number, and the
external force field respectively.  We are using units in which the mass 
density of the fluid is equal to one. Also, the total derivative is defined by

\beq
 {D a \over{Dt}}={\pdr a \over \pdr t} + {\bf u} \cdot \grad a.
\eeq

Pressure can be eliminated by taking the curl of this  equation. This gives
\beq 
{D{\mbf \omega}\over Dt}={\mbf \omega}\cdot\grad {\bf
u}+{R^{-1}\Delta {\mbf \omega}}+ \curl {\bf f}, 
\eeq 
where ${\mbf \omega}=\curl {\bf u}$ is the
vorticity.

Note that for incompressible flow, vorticity determines velocity:
\beq
	u_a(x)=\eps_{abc}\pdr_b\int \omega_c(y) G(x,y)d^3y,
\eeq
or 
\beq
	u_a(x)=\eps_{ab}\pdr_b\int \omega(y) G(x,y)d^2y,
\eeq
in two dimensions. Here \m{G(x,y)} is the Green's function of the Laplace
operator \m{\Delta_x G(x,y)=-\delta(x-y)} with the appropriate boundary conditions
on vorticity. For example, if \m{{\mbf \omega}\to 0} as \m{|x|\to \infty},
\m{G(x,y)=-{1\over 4\pi |{\mbf x}-{\mbf y}|}} in three dimensions and
\m{G(x,y)=-{1\over 2\pi}\log|x-y|} in two dimensions.

If the flow is two dimensional (i.e., \m{\mbf u} is independent of the
\m{z} coordinate), vorticity is orthogonal to the \m{xy}-plane so that 
${\mbf \omega}\cdot\grad {\bf u}=0$. Moreover vorticity  can be thought of as  a
pseudo-scalar. Thus, for two dimensional incompressible inviscid (i.e., $R^{-1}\to 0$) flow
without external forces, vorticity is conserved along flow lines\footnote{For three dimensional flow, vorticity is a vector field.  Inviscid flow without external forces satisfies \m{{\pdr {\mbf \omega}\over \pdr t}+{\cal
L}_{\mbf u}{\mbf\omega}=0}, where the Lie derivative of vector fields is
\m{\cal L} is  given  by 
\m{{\cal L}_{\mbf u}{\mbf\omega}={\mbf u}\cdot\nabla{\mbf
\omega}-{\mbf \omega}\cdot\nabla {\mbf u}}. Thus, vorticiy is
still conserved by the flow.}, with 
\beq
{D { \omega} \over Dt}=0, \;\;\;\;\; { \omega}=\eps_{ab}\pdr_a u_b.
\eeq

By expressing velocity in terms of vorticity we get an integro-differential
equation:
\beq
	{\pdr \omega \over \pdr t}=\pdr_a\omega\eps_{ab}
	\pdr_b\int G(x,y)\omega(y)d^2y.
\eeq

These are analogous to another system studied by Euler: the equations 
for the rigid body given by
\beq
	{dL_a\over dt}=\eps_{abc}L_b A_{cd}L_d,
\eeq 
where \m{A_{ab}} is the inverse of the moment of inertia.
The integral operator on the rhs is analogous to the inverse of  moment of inertia. Thus
vorticity corresponds to angular momentum, the stream function to angular
velocity and the Laplace operator to the moment of inertia.
 The deep reason for this analogy\cite{arnold} is that both systems describe equations for geodesics on a
group manifold. In the case of the rigid body, it is the rotation group
with the metric given by the moment of  inertia tensor; in the case of
hydrodynamics, it is the group of area preserving diffeomorphisms with the
\m{L^2}-metric. In more physical terms, this analogy can be seen in 
the hamiltonian formalism.

Recall that the space of functions on the plane is a Lie algebra, with  
 the Lie bracket given by:
\beq
	[f_1,f_2]=\eps^{ab}\pdr_af_1\pdr_bf_2.
\eeq
Every function \m{f:R^2\to R} corresponds to  an observable of two dimensional
hydrodynamics, \m{\omega_f=\int f(x)\omega(x)d^2x}. The above Lie bracket of
functions suggests a natural postulate of  Poisson bracket for vorticity:
\beq
	\{\omega(x),\omega(y)\}=\eps^{ab}\pdr_b\omega(x)\pdr_a\delta(x-y); 
\eeq
so that, \m{\{\omega_{f_1},\omega_{f_2}\}=\omega_{[f_1,f_2]}}.
 The  natural postulate for the hamiltonian is the total energy of the fluid, 
\beq
H=\half \int {\mbf u}^2 d^2x = {1\over 2} \int G(x,y)  \omega(x) \omega(y) d^2x d^2y.
\eeq

A straightforward calculation shows that
 this hamiltonian with the above Poisson bracket for vorticity indeed 
 gives the  equation of motion of two dimensional hydrodynamics.  

This is analogous to the hamiltonian formalism of rigid body mechanics. The
Poisson bracket of the angular momentum (as measured in the co-moving
coordinates) arises from the Lie algebra of infinitesimal rotations,
\beq
\{L_a,L_b\}=\eps_{abc}L_c.
\eeq
The hamiltonian is the rotational kinetic energy 
\beq
	H=\half A_{ab}L_aL_b,
\eeq
where \m{A} is the inverse of the moment of inertia matrix. Thus, vorticity is
analogous to angular momentum; the Lie algebra of functions on the plane
analogous to the Lie algebra of rotations; and the Green's function of the Laplace
operator analogous to the inverse of the moment of inertia matrix.

The square of the angular momentum \m{L^2=L_aL_a} is a central function (i.e.,
has zero Poisson bracket with all functions of angular momentum). In particular it commutes with the hamiltonian
and hence is conserved. The phase space of rigid body mechanics is the sphere on
which \m{L^2} is constant: this is the `symplectic leaf' of the Poisson algebra. 

Analogously, the moments 
\beq
	Q_k=\int \omega^k(x)d^2x, \quad k=1,2,\cdots 
\eeq
are central functions (hence conserved quantities) in two dimensional
hydrodynamics; the phase space of two-dimensional hydrodynamics is the set
of all vorticity fields with a given set of values of these \m{Q_k}.

 The information in these moments can be packaged 
into  the {\em vorticity distribution function} \m{\rho(\lambda)},
\beq
	\int_{-\infty}^\infty  \rho(\lambda)\lambda^k d\lambda=\int \omega(x)^k d^2x.
\eeq
Geometrically,  $\rho(\lambda)$ is the perimeter of the contour curve of vorticity
where $\omega(x)=\lambda$.  This can be written explicitly as
\beq
\rho(\lambda)=\int \delta(\omega(x)-\lambda) d^2x.
\eeq
The shape of these curves changes with time but not their perimeter.

In addition to these central functions, there may be additional conserved
quantities such as momentum and angular momentum:
\beq
	P_a=\eps_{ab}\int x^b\omega(x)d^2x, \;\;\; \quad L=\half \int x^2 \omega(x)d^2x.
\eeq
But the presence of a boundary can break translational or rotation invariance,
violating these conservation laws.

In spite of the presence of these infinite number of conservation
laws, two dimensional hydrodynamics is far from being integrable: the
phase space on which the \m{Q_k} are constant is still infinite-dimensional.  The crucial  step in a statistical
approach is the correct identification of entropy. This is ultimately defined by
the canonical structure of the phase space: entropy of a macroscopic
configuration is the log of the volume of the phase space corresponding to it. 
We would for example like to determine the volume of the phase space with a given
set of values of moments \m{Q_k}.  However, this is some infinite dimensional
subspace of the space of all vorticity distributions. We need to find  a way to
approximate the phase space by a finite dimensional space: a `regularization' or
`discretization' of the system.  The entropy can then be
determined within this discretized version of the theory and eventually a limit
as the number of degrees of freedom goes to infinity can be taken.

We now describe an elegant discretization\cite{discrete} of two dimensional hydrodynamics that
preserves its symmetries (Lie algebra structure) and conservation laws. We will
see that the problem of determining the entropy of a two dimensional flow can
then be solved using ideas from Random matrix theory due to Wigner.

\section{Regularization}

It will be convenient to assume that the flow satisfies periodic boundary
conditions in the plane. Our final result for entropy will be the same even if
other boundary conditions are assumed, but the intermediate formulas seem
simplest with periodic boundary conditions:
\beq
\omega(x_1+L_1,x_2)=\omega(x_1,x_2),\quad \omega(x_1,x_2+L_2)=\omega(x_1,x_2).
\eeq
We can then Fourier analyze vorticity:
\beq
	\omega_{m_1m_2}=\int_{0}^{L_1}\int_{0}^{L_2}\omega(x)
	e^{-2\pi i{m_1x_1\over L_1}} e^{-2\pi i{m_2x_2\over
	L_2}}d^2x,
\eeq
\beq
	\omega(x)={1\over L_1L_2}\sum_{m_1,m_2=-\infty}^{\infty}\omega_{m_1m_2}e^{2\pi i{m_1x_1\over L_1}} e^{2\pi i{m_2x_2\over
	L_2}}.
\eeq
The hamiltonian and Poisson bracket are, in terms of these variables,
\beq
H=(L_1L_2)^2\sum_{m\neq (0,0)} {1\over m^2}|\omega_{m}|^2,\quad 
\{\omega_{m},\omega_{n}\}=-{2\pi\over L_1L_2}\eps_{ab}m_an_b
\omega_{m+n}.
\eeq
The equations of motion become
\beq
	{d\omega_{p}\over dt}=\sum_{m+n=p}\eps^{ab}m_an_b\left[{1\over
	m^2}-{1\over n^2}\right]\omega_m\omega_n.
\eeq
We are still dealing with a system with an infinite number of degrees of freedom.
At first glance, ignoring  all except  the low momentum modes (i.e., keeping only  \m{|m_1|,|m_2|<N} for some \m{N}) 
looks like a reasonable `regularization' of the problem: we would lose only information at very small length scales.  However, such a naive truncation that simply ignores large \m{m_1,m_2} modes would not
be consistent: for example, the Poisson bracket of two low momentum modes could
be a high momentum mode. We must modify the Poisson brackets so that the Poisson
brackets of the modes we keep form a closed Lie algebra;
 as \m{N\to \infty} this modification must dissappear and we must recover the
 original Lie algebra. Moreover the hamiltonian must be modified so that the
 equations don't mix high and low momentum modes.
 
 A  formalism along these lines is known in the literature\cite{discrete}. The essential idea is to modify
  the coefficients in the Poisson brackets and hamiltonian so that they are periodic with period
  \m{N}; this way the algebra would be closed.  There is a choice that satisfies the Jacobi identity, thus preserving the Lie algebra structure:
\beq
	\{\omega_m,\omega_n\}={1\over \theta}\sin[ \theta(m_1n_2-m_2n_1)]\omega_{m+n\mod
	N},\quad \theta={2\pi\over N}.
\eeq
We keep only a finite number of Fourier modes (we can assume that \m{N}
is odd \m{N=2\nu+1}),
\beq 
 m_1,m_2,n_1,n_2=-\nu,\cdots 0,1,\cdots \nu.
\eeq
To be fair, the hamiltonian is truncated as well:
\beq
	H=\half\sum_{\matrix{m_1,m_2=-\nu\cr \lambda(m)\neq 0}}^{\nu}
	{1\over \lambda(m)}|\omega_m|^2.
\eeq
It is natural to deform the eigenvalues of the Laplacian to 
\m{\lambda(m)=\left\{{N\over
2\pi}\sin\left[{2\pi\over N} m_1\right]\right\}^2+\left\{{N\over
2\pi}\sin\left[{2\pi\over N} m_2\right]\right\}^2}.  This would preserve the
periodicity of the eigenvalue modulo \m{N}.

It is clear that as \m{N\to \infty}, \m{\theta\to 0} and the structure constants of
the Lie algebra tend to the original ones; so does the hamiltonian. 
The Lie algebra we obtain this way is in fact nothing but that of the Unitary
group \m{U(N)}, the Lie algebra of hermitean matrices. Indeed the discretization procedure above has a natural
interpretation in terms of non-commutative geometry; the Lie algebra is the
algebra of derivations of the algebra of functions on a non-commutative torus.
The constants \m{\lambda(m)} in the hamiltonian above are the eigenvalues of the
Laplace operator. However, we will not need this interpretation in what follows. 
Non-commutative geometry plays a deeper role in the three dimensional version of
this theory, which one of us is still developing.

\section{Matrix Formulation}

It will be convenient to make a linear change of variables that will make the
connection with hermitean matrices more explicit. Define \m{N\times N} unitary
matrices \m{U_1,U_2} satisfying
\beq
	U_1U_2=e^{i\theta}U_2U_1.
\eeq
Then, 
\m{U(m)=e^{-{i\over 2}m_1m_2\theta}U_1^{m_1}U_2^{m_2}} satisfies
\m{U^\dag(m)=U(-m)}.
To be specific we choose,
\beq
	U_1=\pmatrix{1&0&\cdots&0\cr
		  0&e^{i\theta}&\cdots&0\cr
		  \cdot&\cdot&\cdots&\cdot\cr
		  \cdot&\cdot&\cdots&e^{i(N-1)\theta}},\quad
       U_2=\pmatrix{0&0&\cdots&0&1\cr
                 1&0&\cdots&0&0\cr
		 \cdot&\cdot&\cdots&\cdot&\cdot\cr
		 0&0&0&\cdots&1&0}.
\eeq
(These matrices can be viewed as the coordinates on a non-commutative torus.)
Now the Fourier coefficients of vorticity can be packaged in the hermitean
matrix
\beq
	\hat\omega =\sum_{m}\omega_{m}U(m).
\eeq
(Recall that \m{\omega(x)} being real implies that
\m{\omega^*_{m}=\omega_{-m}}. This along with \m{U^\dag(m)=U(-m)} implies that 
\m{\hat\omega^\dag=\hat\omega}.)
Now we can verify that 
\beq
\{\hat\omega_{ab},\hat\omega_{cd}\}=i[\delta_{bc}\hat\omega_{ad}-\delta_{da}\hat\omega_{cb}].
\eeq
These are  the well-known commutation relations of the Lie algebra of \m{U(N)}
in the Weyl basis; this establishes the identification of our 
 truncated Lie algebra. Moreover, we can see that the quantities 
 \beq
 	\hat Q_k=\tr \hat \omega^k,\quad  k=0,1,\cdots N-1
 \eeq 
 are central in this algebra: they commute with any function of \m{\hat\omega}.
 In the limit as \m{N\to \infty, \theta\to 0}, these tend to the central
 functions \m{Q_k} we started with. 
 
 The hamiltonian becomes,
 \beq
 	H=\half \hat \omega_{ab}\hat \omega_{cd}G_{abcd}
 \eeq
 for some  tensor \m{G_{abcd}} whose explicit form we do not need. The equation
 of motion are
 \beq
 	{d\hat\omega\over dt}=i[\Omega,\hat\omega]
 \eeq
 where we have \m{\Omega_{ab}=G_{abcd}\hat\omega_{cd}}. Thus time evolution is a
 sequence of unitary transformation so that the traces of powers of
 \m{\hat\omega} are unchanged.

 \section{Formula for Entropy}
 
 Now it is clear that the information contained in the moments
 \m{\hat Q_k=\tr\hat\omega^k=\sum_{j=1}^N\lambda_j^k} is simply the 
 spectrum \m{\{\lambda_1,\cdots ,\lambda_N\}} of the hermitean matrix
 \m{\hat\omega}. The phase space of the system is a vector space of finite
 dimension \m{N^2}. The submanifold of matrices with a fixed
 spectrum \m{\{\lambda_1,\cdots \lambda_N\}} is  compact and has dimension 
 \m{N(N-1)}: 
 it is called the `flag manifold' in algebraic geometry. There is a unique 
 (up to a multiplicative constant) volume form on this submanifold invariant under the action of the unitary group.  (This
 is the volume form induced by the symplectic structure associated to the
 Poisson brackets above.) The volume of microstates with a fixed value of moments \m{\hat Q_k} are given by
 integrating this volume form.  

The volume of this submanifold is well known\cite{mehta}:
  \beq
 	V_N=C_N\prod_{1\leq k< l\leq N}\left[\lambda_k-\lambda_l\right]^2. 
 \eeq
 Here \m{C_N} is a constant whose value we do not need.
 Thus the log of the volume of this manifold becomes
 \beq
 \log V_N=2\sum_{1\leq k< l\leq N}\log|\lambda_k-\lambda_l|+\log C_N.
 \eeq
 In terms of  the density distribution  of eigenvalues
 \beq
 	\rho_N(\lambda)={1\over N}\sum_{k=1}^N\delta(\lambda-\lambda_k),
 \eeq
 this becomes
 \beq
 \log V_N=N^2{\cal P}\int
	\rho_N(\lambda)\rho_N(\lambda')\log|\lambda-\lambda'|d\lambda d\lambda'+\log
	C_N.
 \eeq
 Here \m{{\cal P}\int} denotes the principal value integral:
 \beq
 	{\cal P}\int f(\lambda,\lambda')d\lambda d\lambda'=\lim_{\eps\to
	0^+}\int_{|\lambda-\lambda'|>\eps}f(\lambda,\lambda')d\lambda d\lambda'.
 \eeq
 The advantage of this point of view is that it survives the limit \m{N\to
 \infty}, except for an undetermined additive constant. 
 
 Now, in the limit \m{N\to \infty}, 
 \beq
 	\int \rho_N(\lambda)\lambda^k d\lambda \to {1\over A} \int \rho(\lambda)\lambda^kd\lambda, 
 \eeq
 where \m{\rho(\lambda)=\int \delta(\omega(x)-\lambda)d^2x} as we noted 
 earlier. Also \m{A=\int d^2x} is the total area of the region of the
 flow. Thus we see that \m{\lim_{N\to
 \infty}\rho_N(\lambda)={\bar\rho}(\lambda)}, where 
 \m{{\bar \rho}(\lambda)={1\over A} \rho(\lambda)} is the normalized 
 distribution of vorticity: \m{\int \bar\rho(\lambda)d\lambda=1}. We can
 now get the limiting form of the log of the volume of the submanifold with a
 given set of moments:
 \beq
 	\chi(\omega):=\lim_{N\to \infty}{1\over N^2}\log V_N={\cal P}\int
	\bar\rho(\lambda)\bar\rho(\lambda')\log|\lambda-\lambda'|d\lambda
	d\lambda'.
\eeq
(We drop the undetermined additive constant.) 
We can rewrite this directly in terms of vorticity:
\beq
	\chi(\omega)={\cal P}\int\log|\omega(x)-\omega(y)|{d^2xd^2y\over
	A^2}.
\eeq
This quantity should be viewed as the entropy of a vorticity distribution.

This formula for entropy is one of our main results.  Our 
careful derivation by truncating the Lie algebra gives an entirely different
result from those postulated in earlier analyses\cite{onsager,sommeria,jmiller}, namely,
\beq
	\int \omega(x)\log|\omega(x)|d^2x.
\eeq

\section{Semi-circular distribution}
As an application of this new notion of entropy we describe here the vortex
distribution of maximum entropy with given total vorticity \m{Q_1} 
and enstrophy \m{Q_2}. We need to maximize 
\beq
\chi(\rho)={\cal P}\int
\bar\rho(\lambda)\bar\rho(\lambda')\log|\lambda-\lambda'|d\lambda d\lambda'
\eeq
keeping the quantities \m{\int \bar\rho(\lambda)\lambda^k d\lambda} for
\m{k=0,1,2}  fixed. Introducing Lagrange multipliers \m{c_0,c_1,c_2} for the constrained variational problem,
we get 
 \beq
 	2{\cal P}\int
	\bar\rho(\lambda')\log|\lambda-\lambda'|d\lambda'=c_0+c_1\lambda+c_2\lambda^2. 
 \eeq
 Differentating this to eliminate \m{c_0} gives the singular integral equation
 \beq
 	2{\cal P}\int {\bar\rho(\lambda')\over
	\lambda-\lambda'}d\lambda'=c_1+2c_2\lambda.
 \eeq
 The equation is well known in the theory of random matrices;
 the solution is the `semi-circular' distribution of Wigner\cite{mehta}
 \beq
 	\bar\rho(\lambda)={1\over 2\pi \sigma^2} \theta\left(|\lambda-\bar Q_1|\leq
	2\sigma\right)\surd\left[4\sigma^2-(\lambda-\bar Q_1)^2\right].
 \eeq
\begin{figure}[htbp] 
\vspace*{13pt}
\centerline{\psfig{file=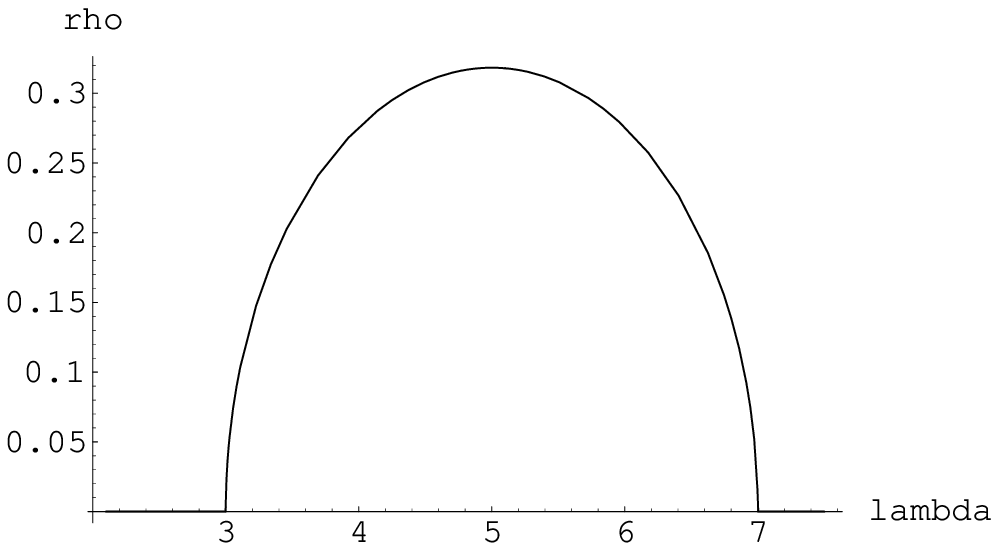}} 
\vspace*{13pt}
\fcaption{Vorticity distribution function of maximum
entropy with  
\m{{\bar Q_1}=5.0, {\sigma}=1.0.}}
\end{figure}

 Here \m{{\bar Q}_1={1\over A}\int \omega(x)d^2x} is the mean value of
 vorticity.
 The constant \m{\sigma} is the standard deviation of the vorticity
 distribution and is determined by the enstrophy per unit area \m{{\bar
 Q}_2={1\over A}\int \omega(x)^2d^2x}:
 \beq
 \bar Q_2={\bar Q}_1^2+\sigma^2.
 \eeq

\section{Vorticity Distribution in a Cylindrically Symmetric Fluid}
As an applicaton of our ideas we determine the maximum entropy configuration of
an axially symmetric fluid trapped between two concentric circles of radii
\m{a_1<a_2}.  The vorticity is a function only of the distance
from the origin in the plane. 

With axial symmetry, {\em any} function \m{\omega(r)} of the radial distance alone will be a
static solution of the Euler equations. We will seek the answer to the
following question: {\em among all cylindrically symmetric functions 
with given mean vorticity and enstropy, which one
has the largest entropy?} One should expect that a system subject to many
perturbations will eventually settle down to this distribution. 

Now, the vortex contours of an axisymmetric vorticity distribution are
concentric circles. We should expect that the vorticity is monotonic with
distance in the stable configuration. The vortex distribution is given by
\beq
\rho(\lambda)=\int \delta(\rho(r)-\lambda)d^2x=2\pi\int_0^\infty
\delta(\omega(r)-\lambda)rdr.
\eeq
Let us introduce the variable \m{u=r^2}. Rather than think of the vorticity
\m{\omega}  as a
function of \m{r} (or \m{u}), let us try to determine its inverse function;
that is, \m{u} as a function of
\m{\omega}. Then we see that 
\beq
\rho(\lambda)=\pi \left|{du\over d\lambda}\right|.
\eeq
From the last section we know \m{\rho(\lambda)=A\bar\rho(\lambda)} to be
the semi-circular distribution.
 We can solve the above
differential equation to determine the corresponding vorticity profile in
parametric form:
\beq
\omega=2\sigma\sin\phi+\bar Q_1,\quad r^2=\half[a_1^2+a_2^2]\pm
[a_2^2-a_1^2]{1\over \pi}\left[\phi+{1\over
2}\sin\left(2\phi\right)\right],
\eeq
with the parameter \m{\phi} taking the range of values
\beq 
 -{\pi\over 2}\leq \phi\leq {\pi\over2}.
\eeq
There are two possible solutions with equal entropy: one with vorticity increasing with
distance and the other with it decreasing. In most situations we would expect 
the vorticity to decrease with radial distance. We give a sample plot of this
distribution below.

\begin{figure}[htbp] 
\vspace*{13pt}
\centerline{\psfig{file=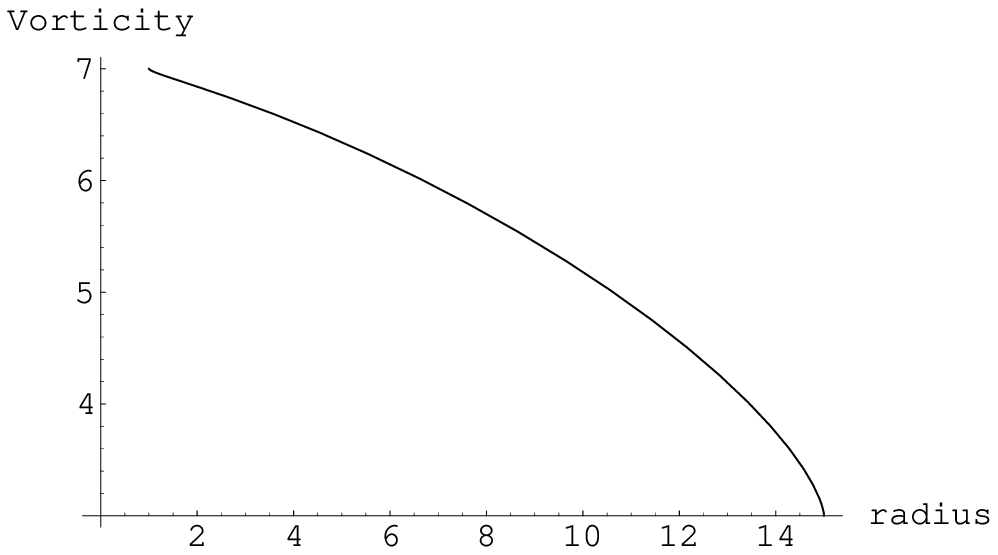}} 
\vspace*{13pt}
\fcaption{Vorticity field  of maximum entropy 
within radii   \m{a_1=1.0,
a_2=15},  mean vorticity \m{{\bar Q_1}=5.0}, and variance \m{\sigma^2}=1.0.} 
\end{figure}

This could describe a vortex such as a hurricane, Jupiter's red spot or a tornado, to a reasonable approximation. A comparison with experimental measurements would be interesting.

\nonumsection{Acknowledgments}
\noindent
The work of SVI was supported by the Dr. Nuala McGann Drescher Award funded by the State of New York/UUP Affirmative Action/Diversity Committee.

\nonumsection{References}


\noindent

\begin{thebibliography}{000}


\bibitem{dezhe}
Dezhe Z. Jin and Daniel H.E. Dubin, Phys. Rev. Lett. {\bibbf 84}, 1443 (2000); D.Z. Jin and D.H.E. Dubin, Phys. Rev. Lett. {\bibbf 80}, 4434 (1998).



\bibitem{onsager}
L. Onsager, ``{\bibit Statistical Hydrodynamics}," Nuovo. Cim. Suppl. {\bibbf 6}, 279 (1949).

\bibitem{sommeria}
R. Robert and J. Sommeria, ``{\bibit Statistical Equilibrium States for Two-dimensional Flows}," J. Fluid Mech. {\bibbf 229}, 291 (1991).

\bibitem{jmiller}
Jonathan Miller, ``{\bibit Statistical Mechanics of Euler Equations in Two Dimensions},"
Phys. Rev. Lett. {\bibbf 65}, 2137 (1990); J. Miller, M.C. Cross and P.B. Weizmann, ``{\bibit Statistical Mechanics, Euler's Equation, and Jupiter's Red Spot}," Phys. Rev. A {\bibbf 45}, 2328 (1992).

\bibitem{mehta}
M.L. Mehta, ``{\bibit Random Matrices}," Academic Press (1991).

\bibitem{polyakov} A. Polyakov, Nucl. Phys. B {\bibbf 396}, 367
(1993); hep-th/9212145.

\bibitem{discrete}J. Hoppe, Int. J. Mod. Phys. A {\bibbf 4}, 5235
(1989); D. B. Fairlie and C. K. Zachos, Phys. Lett. B {\bibbf
218}, 203 (1989); D. B. Fairlie, P. Fletcher and C. K. Zachos, J. Math. Phys. {\bibbf 31}, 1088 (1990);
J.S. Dowker and A. Wolski, ``{\bibit Finite Model of Two-dimensional Ideal Hydrodynamics}," Phys. Rev. A {\bibbf 46}, 6417 (1992).

\bibitem{arnold}
V.I. Arnol'd, ``{\bibit Sur La Geometrie differentielle des groupes de Lie de dimension infinie et ses applications a l'hydrodynamique des fluides parfaits}," Ann. Inst. Fourier {\bibbf 16}, 319 (1966).

\end{thebibliography}
\end{document}